\documentclass[10pt, conference, compsocconf]{IEEEtran}

\ifCLASSINFOpdf
\else
\fi
\usepackage[colorinlistoftodos,prependcaption,textsize=small]{todonotes}
\usepackage{blindtext}
\presetkeys%
    {todonotes}%
    {inline,backgroundcolor=yellow}{}
\usepackage{url}

\graphicspath{{.}{images/}}
\DeclareGraphicsExtensions{.pdf}

\begin{document}
\title{Smart Shires: The Revenge of Countrysides (Extended Version)\footnotemark}


\author{\IEEEauthorblockN{Stefano Ferretti, Gabriele D'Angelo}
\IEEEauthorblockA{Department of Computer Science and Engineering, University of Bologna\\
Bologna, Italy\\
\{s.ferretti, g.dangelo\}@unibo.it}
}


%Telephone: (800) 555--1212, Fax: (888) 555--1212}


\maketitle

\footnotetext{The publisher version of this paper is available at \url{http://dx.doi.org/10.1109/ISCC.2016.7543827}.
\textbf{{\color{red}Please cite this paper as: ``Stefano Ferretti, Gabriele D'Angelo. Smart Shires: The Revenge of Countrysides. Proceedings of the IEEE Symposium on Computers and Communication (ISCC 2016), Messina (Italy), 2016.''}}}

\begin{abstract}
This paper discusses the need to devise novel strategies to create smart services specifically designed to non-metropolitan areas, i.e.~countrysides. These solutions must be viable, cheap an should take into consideration the different nature of countrysides, that cannot afford the deployment of services designed for smart cities. These solutions would have an important social impact for people leaving in these countrysides, and might slow down the constant migration of citizens towards metropolis. In this work, we focus on communication technologies and practical technological/software distributed architectures. An important aspect for the real deployment of these ``smart shires'' is their simulation. We show that ``priority-based broadcast'' schemes over ad-hoc networks can represent an effective communication substrate to be used in a software middleware promoting the creation of applications for smart shire scenarios.
\end{abstract}


% \fi
\IEEEpeerreviewmaketitle

\section{Introduction}

In last years, recent advances in the ICT domain were focused mainly on “smart cities”, i.e. a set of (in certain cases technological) strategies aiming at improving and optimizing services offered to citizens. In most cases, all these services are devoted to dense metropolitan areas. Pervasive computing and mobile services represent important technologies that may help and guide the citizens in their daily activities. These projects have a big impact and may add several benefits to the society. In the long-term, all these efforts might have two important social effects: on one hand they would improve the life of the citizen; but on the other hand, they might even further push other citizens to leave the countrysides and rural areas towards metropolitan areas. As a confirmation of this claim, some seminal papers on smart cities assert that ``over the next three decades, seventy percent of the global population will live in cities'', e.g.,~\cite{perez}.

As a matter of fact, the possibility to offering services for territorial districts with low population density is an almost ignored problem. We are not referring here to the well-known ``digital divide'', a gap that has been widely recognized and that, in certain areas of more civilized countries (e.g., Europe), is being solved, more or less rapidly. Several (non-metropolitan) territories, composed of small towns, can offer modern (yet, by now, traditional) networking infrastructures. However, surrounding areas are commonly not covered by further communication services, besides network connectivity. Indeed, to follow the smart cities trend, connectivity is a necessary but not sufficient feature.

Non-metropolitan areas can be very different and various depending on the specific geographical location we are considering, e.g.~European countrysides are very different to those that can be found in the Americas or in Asian countries. Anyhow, while very diverse, these areas share a lack of innovative solutions that optimize the use of resources.  We claim the need to focus on these areas, that require smart management solutions.

There is plenty of country depopulated territories which do have a huge and underestimated potential, due to the beauty of their sceneries, the healthy lifestyle, the possible (but not exploited) tourist potential.
If information and communication are considered the keys to the intelligent cities of tomorrow, the same must hold for decentralized areas. 
The solution is not, for instance, adding wireless antennas by itself. Neither, it is not possible implementing (costly) smart cities services to make them work in a country territory, due to the very different economic circumstances that would make these services not feasible in such context. There is the need for innovative, self-configuring and cheap solutions, possibly not strictly dependent to the presence of a classic networking infrastructure. Rather, such solutions should be opportunistic, in the sense that schemes must be available, which allow mobile nodes exploiting network infrastructures, when possible, and alternative connectivity solutions in other circumstances. This would create a middleware platform on top of which novel smart services might be deployed.

A ``smart shire'' is a novel view of a geographical space able to manage resources (natural, human, equipment, buildings and infrastructure) in a way that is sustainable and not harmful to the environment. Examples of technologies that might compose the substrate for developing of a smart shire platform are: multihoming mobile services, mobile ad-hoc networks, opportunistic networks, peer-to-peer and cloud (or alternatively, fog) computing systems.

A smart middleware can be built that leverages these services; such a middleware would simplify the development of new services and the
integration of legacy technologies into new ones~\cite{Atzori:2010}. On top of the middleware, smart services will be deployed whose application domains include information dissemination, tourist services in areas where a communication infrastructure is not present, infomobility, weather services, pollution in rural areas (e.g., dumps), sustainable (sub)urban environment, healthy services for ageing population, continuous care, emergency response, smart renewable energy.

The development of a smart shire would not only impact on the willing of population to migrate to the city. Countrysides and cities are strongly interconnected. Rural areas are providers of goods for cities (think at products from the agroindustry, for example) and thus, optimizing services for countrysides would have a positive impact for smart cities~\cite{robledo}. In substance, the deployment of innovative, cheap smart services for countrysides would allow interconnecting these novel services with those offered in smart cities, leading to the formation of a connected smart territory, seen as a complex system where resources are viably exploited, leading to a sustainable environment.

An important methodology to preliminarily assess the viability of large scale solutions in wide areas, such as countrysides, is simulation. In this case, simulation must be carefully handled in order to create reasonably accurate models that can scale in terms of modeled entities and granularity of events. Indeed, even a small size smart shire will be composed by thousands of (possibly) interconnected sensors and devices. Thus, scalability is the main requirement to consider. Scalability would allow considering smart shires that are connected to smart cities, thus creating a whole smart territory. To this aim, probably the best approach accounts to the use of discrete-event simulation combined with an agent-based model. 

With this aim, we have developed a Smart Shire Simulator ($S^3$), an agent based model built on top of GAIA/ART\`IS~\cite{Bononi200452}, that is a simulation middleware enabling the seamless sequential/parallel/distributed execution of large scale simulation runs. Using $S^3$, we study the effectiveness of a ``priority-based broadcast'' dissemination scheme employed over ad-hoc networks, where communication among nodes is made possible through direct transmissions among near devices. Results show that dissemination schemes (when coupled with caching schemes) can represent an effective communication substrate to be used in a software middleware promoting the creation of applications for smart shire scenarios.

The remainder of this paper is organized as follows. 
Section \ref{sec:usecase} describes some killer applications that can be deployed in smart shires.
Section \ref{sec:archi} describes the enabling technologies for the development of a middleware platform for smart shires.
Section \ref{sec:simulation} discusses the need for using simulation as a tool to test and assess viable solutions to be deployed in large scale areas. 
Section \ref{sec:eval} shows the results of a preliminary performance evaluation on $S^3$.
Finally, Section \ref{sec:conc} provides some concluding remarks.

%%%%%%%%%%%%%%%%%%%%%%%%%%%%%%%%%%%%%%%%%%%%%%
\section{Killer Applications}\label{sec:usecase}

Countrysides cannot evolve at the same pace of smart cities. Due to the existing technological gaps, such territories need some novel, smart, tailored solutions. Services might allow federating several neighbor municipalities/towns, private or public organization to form a critical mass of offered services and potential users. In this scenario, wireless networks, with specific focus on ad-hoc, 5G Device-to-Device (D2D) communications, multihoming, have great significance in the physical infrastructure of a smart territory.

Services to citizens might be improved Internet access, access to information on the municipalities or organizations using  wireless connected information panels and kiosks, apps promoting citizen participation in general, web portals, georeferenced information for a multitude of user applications (e.g., for tourists). A main use case for smart territories in general is that of proximity-based applications, where devices detect their proximity and subsequently trigger different services, such as proximity-based social networking, gaming, advertisements for by-passers, local exchange of information, smart communication between vehicles, network coverage extension, etc. Other applications include traffic control, alert systems, security and public safety support, where devices provide at least local connectivity even in case of damage to the radio infrastructure. Finally, services related to the production chain in rural environments relate to smart water, smart parks, sportsmen care, smart metering, smart agriculture, smart animal farming.

Services to municipalities and organizations might be system network-based video surveillance (their implementation using traditional infrastructure-based architectures would be harder, and in proportion more costly, than in cities), smart traffic management systems, traffic light control, data collection through environmental sensors for monitoring resources and facilities, smart eHealth,
security and emergencies. In order to being ready for critical events, such as natural disasters (e.g., earthquakes, floods, fires) sensor networks can be proficiently deployed in the territories (e.g., riverbanks, woods), to be used used in smart services. Gathered information would allow municipalities modeling (and simulating) the behavior of ecosystems, in order to improve decision-making to manage the whole territory. Other applications might refer to the optimization of the supply chain processes in the rural territory, in conjunction with urban regions in the smart territory.

%%%%%%%%%%%%%%%%%%%%%%%%%%%%%%%%%%%%%%%%%%%%%%
\section{System Architecture}\label{sec:archi}

\begin{figure}[th]
\centering
\includegraphics[width=\linewidth]{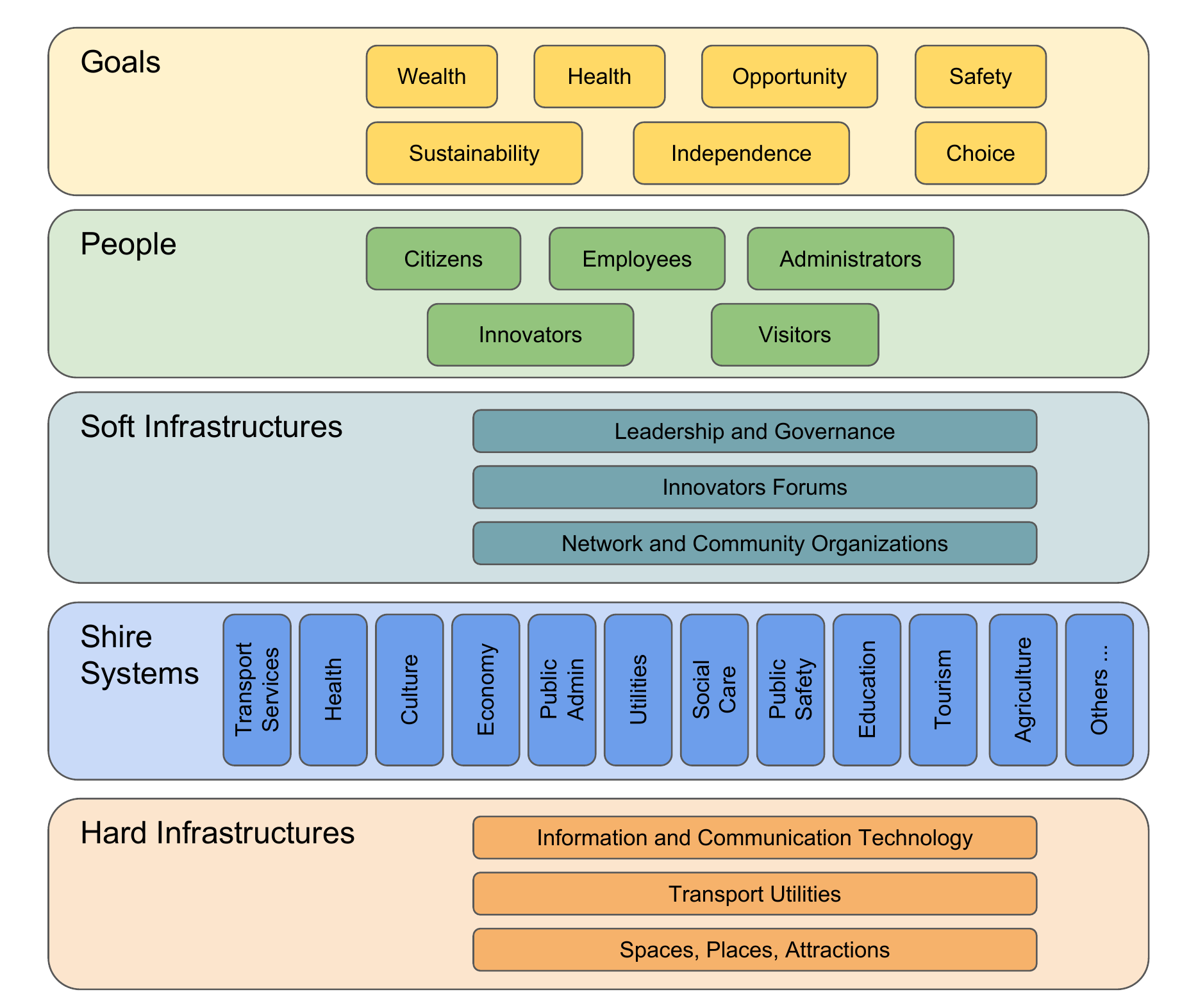}
\caption{Components of a smart shire architecture.}
\label{fig:archi}
\end{figure}

In order to design the architecture and systems devoted to the development of a smart shire, it is important to understand the environment and other elements with which they will interact. Taking inspiration from~\cite{smartCityArchi}, the system and architecture of a smart shire can be described through the context, environments, actors and elements, as reported in Figure \ref{fig:archi}. Every initiative to create a smart territory in general is based on a set of social goals. A not complete set of goals are those reported in the figure: offering improved wealth, health, opportunity, safety, sustainability, independence, choice. These goals should be offered to people, and can be achieved only if people participate to the services. People is a generic term that comprises citizens, (public and private) employees, administrators, innovators and visitors (in several countrysides, tourism can be an important sector to drive and revitalize economic growth). In the process of understanding how communities and individuals might interact with a smart shire, elements of ``soft infrastructure'' are created, i.e.~organizations and interest groups who support shire communities, such as governance entities, groups of innovators, network and community organizations~\cite{smartCityArchi}. Clearly enough, shire systems are significant elements to be the optimized and that will be used by smart services. These systems comprise to the whole spectrum of social services involving the life of citizens. Finally, hard infrastructures represent the physical infrastructures, technologies and platforms that are used to create a middleware able to support application services. Those technologies and platforms include networks such as 4G and broadband, but also ``infrastructure-less'' networking approaches (e.g., MANETs, Internet of Things); communication tools such as social media and audio-video conferencing; computational resources such as cloud or fog computing; information repositories to support open data. Technological infrastructures and platforms will manage data related to the ``traditional'' services and facilities available in the territory.

\begin{figure}[th]
\centering
\includegraphics[width=\linewidth]{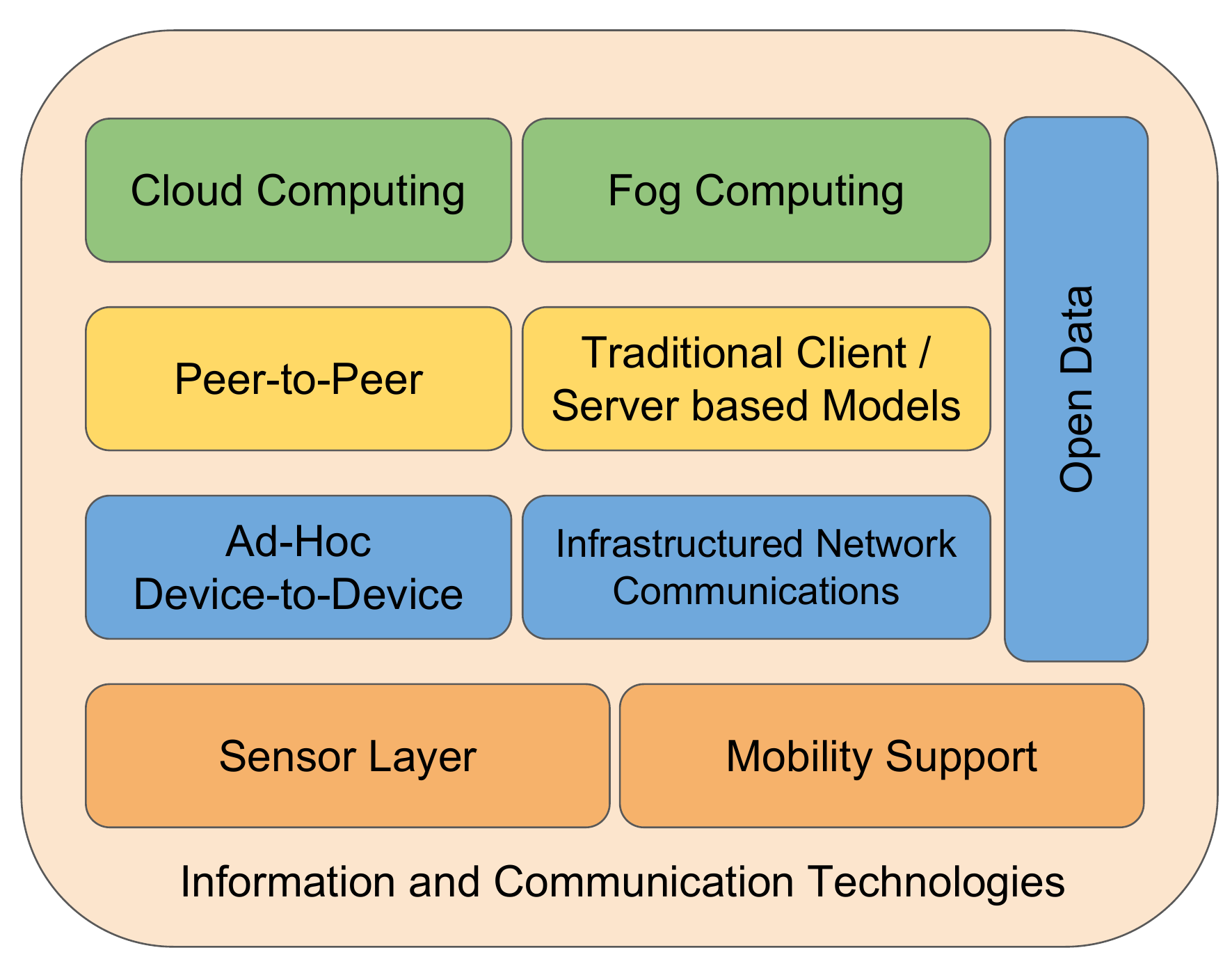}
\caption{Information and communication technologies.}
\label{fig:tech}
\end{figure}

This paper focuses on the communication and technological aspects related to the architecture described above. The set of key technologies to be used to build a middleware platform for smart shires are those reported in Figure \ref{fig:tech}. In the following, we describe the main innovative communication technologies that are into the smart shire picture.

\subsection{Sensing as a Service}
Smart shires should make use of crowd-sensed and crowd-sourced data to generate information to be used within services. It will be important to resort to any possible information coming from the cloud of things available in the territory. An interesting approach is that of considering such cloud of things using the Sensing as a Service (SaaS) model, based on Internet of Things infrastructure~\cite{Khan:2012,Lea:2014,PereraZCG14}. Such SaaS model would enable sharing and reusing of sensor data to create complex services.

Sensors are relatively cheap in terms of costs. Thus, their deployment in a countryside is feasible. Problems may arise to interconnect these sensors to form a sensor network and to make them communicate with the intelligent service placed in the Internet. This requires the use of smart services employing D2D and multihop/multipath communications.

It makes sense to consider the scenario of various heterogeneous devices interconnected to each other and to exploit these interconnections to create novel services~\cite{Petrolo:2014}. Data sensed by the sensors' devices are disseminated and collected by an information processing system, managed as open data within the middleware, to be used by applications. Such a middleware should enable a context-aware data distribution, i.e.~it should be able to distribute context data to interested entities~\cite{Bellavista:2012}.

\subsection{Open data, Crowdsourcing, Crowdsensing}
There is a need to to provide an open data platform to promote collaboration and data sharing among business companies, municipalities and citizens.

Today, there are several examples of online databases where citizens can register their sensors to the aim of gathering data from the environment to be collaboratively exploited by diverse applications. Xively and Wikisensing are two prominent examples of systems offering a set of APIs for the retrieval of open data~\cite{boyle}. Such open data can be exploited in variety of ways. An example (in the smart cities context) is the app developed by Birmingham Parkopedia, which offers parking information in more than 6 thousand cities in 52 different countries. Clearly enough, such open data have a huge potential in smart shire environments, that typically lack the availability of digital data.

\subsection{Multihoming Mobility Support}
Under intermittent network connectivity, enabling interaction between smart objects and mobile users in the IoT becomes a real challenge~\cite{Wirtz:2014}. Smart territories should be able to exploit pervasive solutions, where Internet of Things, wireless sensor networks and ubiquitous computing are merged in a single platform. At the user-side, mobile users can be connected to several wireless access technologies simultaneously and can move seamlessly between them~\cite{Ferretti2016390,GhiniJSS}. This means ensuring that if a mobile node changes its point of attachment to the Internet, while in movement, no communication interruptions are perceived at the application level, and if such interruptions occur, they do not significantly degrade the Quality of Service delivered at the application level. While throughput remains a major goal of system design, the main concern of mobility architectures is how to best manage situations where a mobile node changes network~\cite{GhiniJSS}. To this aim, several solutions are available in the literature~\cite{Budzisz:2012,Ferretti2016390,GhiniJSS,Paasch:2012}.

\subsection{Device-to-Device, Ad-Hoc Communication, Dissemination Strategies}
The multitude of data generated by sensors, users, public and private organizations, to be made available for a plethora of possible services requires a careful management and dissemination of data in the middleware. The standard client/server approach might be useful in certain contexts, but other dissemination strategies have to be made available.

Example relates to D2D communication strategies in 5G enabling technologies, and MANETs~\cite{andrews}. Such schemes enable users communicating directly with other users without or with partial involvement of the network infrastructure. The most prominent example among these schemes is the use of multi-hop relaying approaches, where nodes cooperatively relay messages to reach a destination. Another approach is proactive content caching at the edge, where peak traffic demands are reduced by proactively serving predictable user demands via caching at base stations and users' devices~\cite{bastug}.

At a different, higher level of abstraction, also more general approaches exploited in peer-to-peer systems should be made available, such as public/subscribe approaches and gossip-style dissemination schemes~\cite{D'Angelo:2009,Ferretti20131631}.

Depending on the application scenario a message might have to be sent from a device towards a certain physical area of the territory (e.g., the device is looking for a sensor to understand weather conditions) or towards a specific destination (as in classic MANET/VANET scenarios), or the message ``is looking for an access point'', trying to reach an Internet host or service. In other cases, applications might require a broadcast of a message; this is the case for alert messages, critical situations, or more simply general information or advertisements. To this aim, an efficient broadcast scheme might be employed that spread messages across devices, trying to avoid transmissions' collisions. The broadcast might contain aggregate contents (marshalling), in order to optimize transmissions. This provides a useful scheme to be employed at higher levels and disseminate requests and contents.

\subsection{Cloud and Fog Computing}
In the smart cities domain, cloud computing represents a main technology for the deployment of smart services at a large scale.
The tremendous amount of data produced by the Internet of Things must be managed by powerful and reliable distributed computation systems. This remains valid also for smart shires. 

However, the need for cheap solutions, geo-referenced services, mobility and locality aware solutions make fog computing (also referred as edge computing, or cloudlet) an interesting alternative option. Similarly to what happens to citizens of rural regions, often considered to be placed at the edge of the country, fog computing moves computations from datacenters to the edges of the network, and resorts to a collaborative multitude of end-users to carry out distributed services.

%%%%%%%%%%%%%%%%%%%%%%%%%%%%%%%%%%%%%%%%%%%%%%
\section{The Need for Simulation}\label{sec:simulation}

The design and implementation of smart shires must be supported by simulation. Due to their size and complexity, the simulation tools are both needed for the design of the architecture and as part of the smart shires system. In other terms, we expect to be able to validate the proposed architecture, before its deployment, using a simulation model. Moreover, a model of the smart shire (updated in real-time) must be run to support the management decisions.

There are many requirements that must be satisfied by the simulation tool of choice. Above all, the scalability in terms of modeled entities and granularity of events. In our view, even a small size smart shire will be composed by thousands of interconnected devices. Many of them will be mobile and each with very specific behavior and technical characteristics. For this reason, scalability is the main requirement that we will investigate in the following of this paper. Moreover, in our view, a smart shire needs to be considered as a whole with the surroundings smart cities. That is, as an integrated system and with a holistic approach. This aspect further increases the need of scalable simulators. Another requirement is that the simulator should be able to run in (almost) real-time average size model instances. This is fundamental for a proactive approach (e.g.~simulation in the loop) and to perform ``what-if analysis'' during the management of the deployed architecture. Finally, in most cases a multi-level simulation would be required. In fact, running the whole model at the highest level of detail is unfeasible. A better approach would be to bind different simulators together, each one running at its appropriate level of detail and with specific characteristics of the domain to be simulated (e.g.~mobility models, wireless/wired communications and so on).

Considering the characteristics of the model to be simulated and the requirements described above, in our view the best approach is to use a discrete-event simulation engine coupled with an agent-based smart shire model.

The tool under developed is called Smart Shire Simulator ($S^3$) and is based on the GAIA/ART\`IS simulation middleware~\cite{gda-simpat-2014,pads}. ART\`IS is a simulation middleware that permits the seamless sequential/parallel/distributed execution of large scale simulation runs using different communication approaches (e.g.~shared memory, TCP/IP, MPI) and synchronization methods (e.g.~time-stepped, conservative, optimistic). The GAIA framework has a double tasks: i) to ease the development of simulation models with high level application program interfaces; ii) to implement communication and computational load-balancing strategies, that are based on the adaptive partitioning of the simulation model, with the aim to reduce the simulation execution time.

In the current version, $S^3$ implements a prototype model with a limited set of functionalities that is being used to support the design of the proposed architecture. The many elements composing the smart shire are represented as a set of interacting entities. Some entities are static (e.g.~sensors, traffic lights and road signs) while the others (e.g.~cars and smart-phones) follow specific mobility models. All the entities in the simulated model are equipped with a wireless device. The interaction among entities uses a ``Priority-based Broadcast'' (PbB) strategy that is based on the well-studied probabilistic broadcast strategy~\cite{1200529}. In PbB, every messages that is generated by a node is broadcast to all the nodes that are in proximity of the sender. The message contains a Time-To-Live (TTL) to limit its lifespan and the forwarding is based on two conditions. The first is a probabilistic evaluation (i.e.~probabilistic broadcast) while the second is based on the distance between sender and receiver. In fact, to limit the number of forwarded messages, there is a message forward only if the distance between the nodes is larger than a given threshold. Under the implementation viewpoint, this can be done using a positioning system (e.g.~GPS) if available. Otherwise, the network signal level associated to each received message is used.

In the next section, an initial evaluation of the $S^3$ scalability is presented.

%%%%%%%%%%%%%%%%%%%%%%%%%%%%%%%%%%%%%%%%%%%%%%
\section{Model Description and Performance Evaluation}\label{sec:eval}

In this performance evaluation of $S^3$, a bidimensional toroidal space (with no obstacles) is populated by a given number of devices (i.e.~Simulated Entities, SE). A part of the SEs follows a Random Waypoint (RWP)~\cite{rwp} mobility model while the others are static. The interaction among SEs is based on proximity and implements the multi-hop dissemination protocol described above. The main parameters used in this model are reported in Table~\ref{table:model}. It is clear that the model parameters are strictly dependent on the specific scenario characterized by the geographical and architectural issues of each specific smart shire deployment. This confirms that a simulation based approach is needed to support the design of the architecture and for the appropriate tuning of the runtime parameters. As said before, in this paper we are interested in an overall validation of the proposed approach and of the simulation tool.

\begin{table}[h]
\begin{center}
	\begin{tabular}{ | l | p{4cm} |}
	\hline
  \textbf{Model parameter} & \textbf{Description/Value} \\ \hline
	Number of $\#SEs$ & [1000, 32000] \\ \hline
  Mobility of $\#SEs$ & 50\% Random WayPoint (RWP)\newline 50\% static  \\ \hline
  Speed of RWP & Uniform in the range [1,14] spaceunits/timestep\\ \hline
  Sleep time of RWP & 0 (disabled)\\ \hline
  Interaction range & 250 spaceunits\\ \hline
  Density of $\#SEs$ & 1 node every 10000 $spaceunits^2$\\ \hline
  Forwarding range & $>200$ spaceunits\\ \hline  
  Simulated time & 900 timeunits \\ \hline
  Simulation granularity & 1 timestep = 1 timeunit \\ \hline
  Time-To-Live (TTL) & 4 hops \\ \hline
  Dissemination probability (gossip) & 0.6 \\ \hline
  \end{tabular}
\end{center}
\caption{Simulation model parameters.}
\label{table:model}
\end{table}

All the results reported in this section are averages of multiple independent runs. In all cases, the confidence intervals have been calculated but not reported for readability. The execution platform used in this performance evaluation is a DELL R620 with 2 CPUs and 128 GB of RAM. Each CPU is a Xeon E-2640v2, 2 GHz, 8 physical cores. Each CPU core has Hyper-Threading enabled and therefore the total number of logical cores is 32. The computer is equipped with Ubuntu 14.04.3 LTS and GAIA/ART\`IS version 2.1.0. $S^3$ will be freely available as source code in the next release of GAIA/ART\`IS~\cite{pads}.

\begin{figure}[h]
\centering
\includegraphics[width=8.5cm]{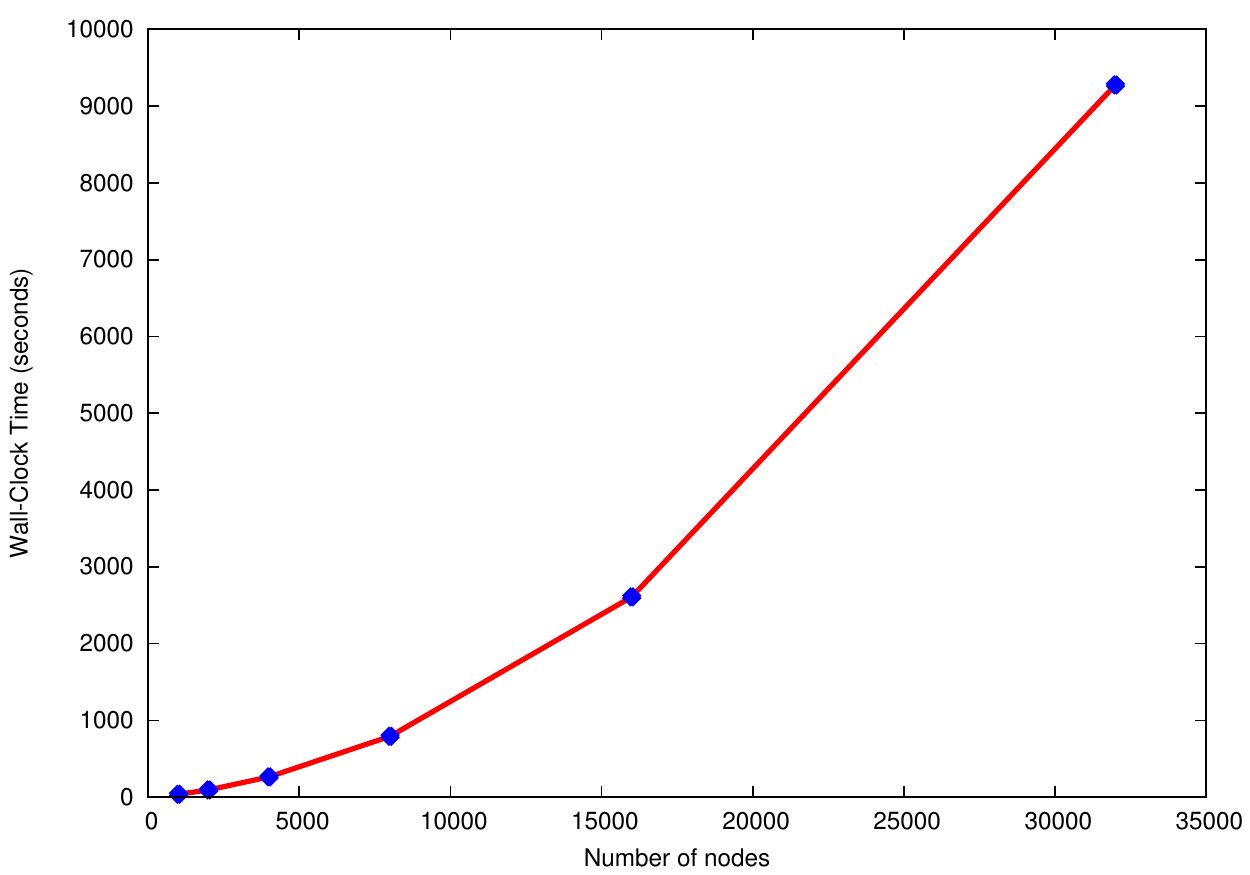}
\caption{Scalability evaluation: increasing number of SEs, sequential (\#CPU core=1) simulator.}
\label{fig_wct_seq}
\end{figure}

In the first experiment, the scalability of the simulator has been evaluated in a sequential setup (that is, 1 CPU core is used). In Figure~\ref{fig_wct_seq} is reported the Wall-Clock Time (WCT) to complete a single simulation run. The number of SEs has been set in the range [1000, 32000]. As expected, the simulator scalability is strongly affected by the number of SEs.

\begin{figure}[h]
\centering
\includegraphics[width=8.5cm]{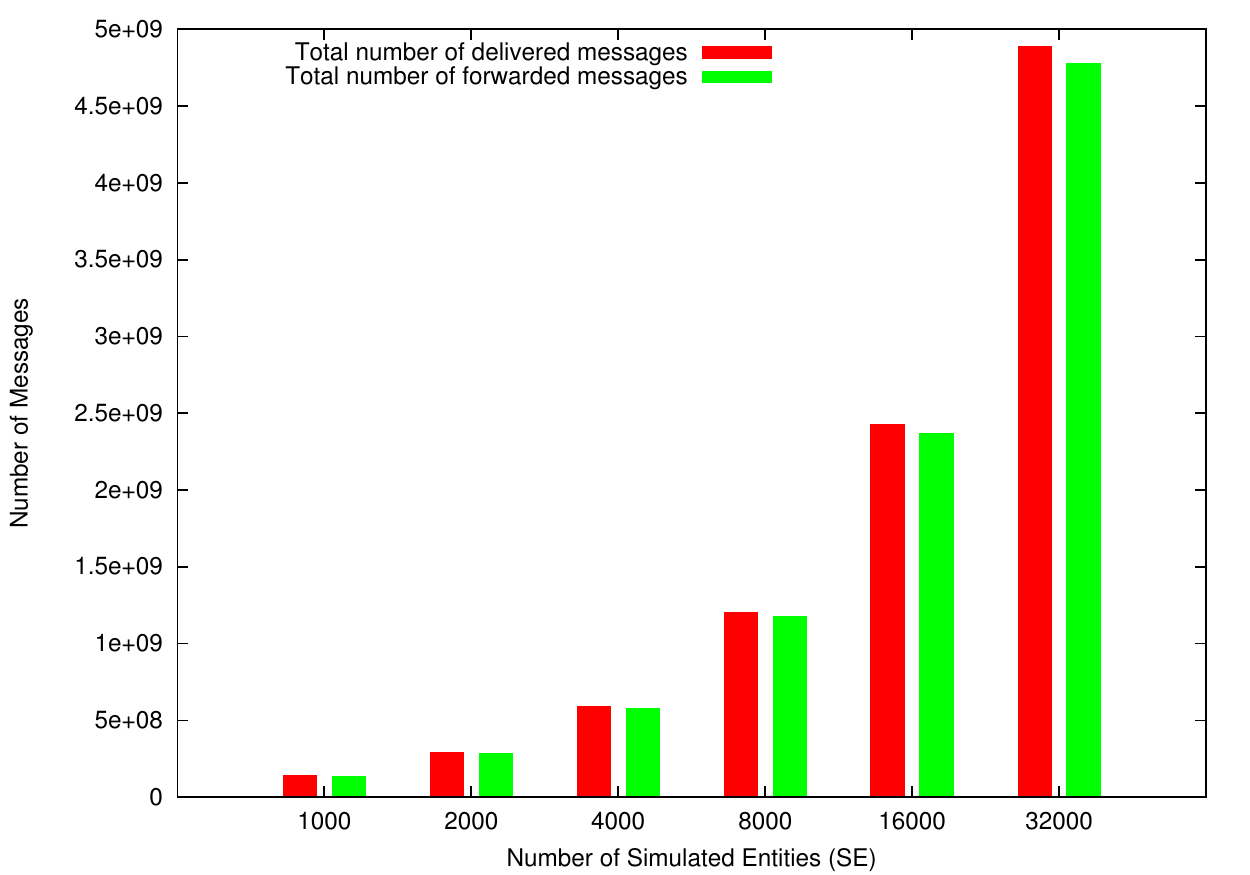}
\caption{Number of delivered messages in a single simulation run: increasing number of SEs, total delivered messages vs. forwarded messages.}
\label{fig_messages}
\end{figure}

Figure~\ref{fig_messages} reports the total number of messages (that is, interactions among devices) measured for each specific scenario. In the current setup, even a limited number of SEs generates a very high number of delivered messages. More specifically, most of them are forwarded messages. This means that a large part of the network traffic is overhead, since each new message is forwarded many times. This severely limits the scalability of the simulation model and may represents a problem for the smart shire architecture. In future revisions of the architecture, appropriate message caching strategies must be considered.

\begin{figure}[h]
\centering
\includegraphics[width=8.5cm]{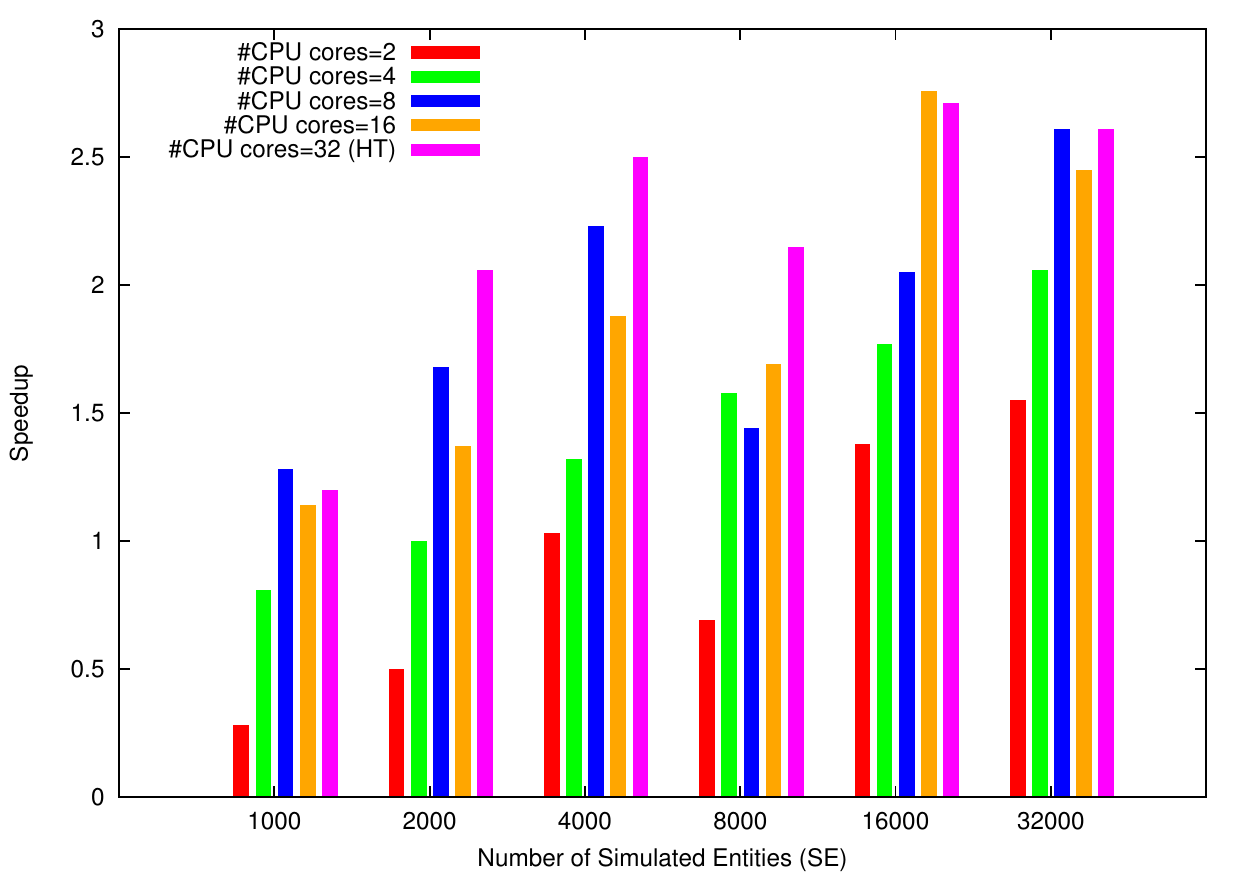}
\caption{Speedup of parallel simulation setup (different number of CPU cores) with an increasing number of SEs.}
\label{fig_speedupoff}
\end{figure}

As seen above, an approach based on a sequential simulator is not adequate due to its scalability constraints. This is the main reason to implement a parallel simulation approach~\cite{gda-simpat-2014} that is able to use all the available CPU cores in the execution architecture. This means that the SEs in the smart shire model have to be partitioned among the CPU cores and the interactions among different SEs have to be delivered using a message passing approach. In Figure~\ref{fig_speedupoff} is reported the speedup (defined as: the ratio between the WCT of the sequential simulation and the WCT obtained by the parallel execution) that can be obtained by GAIA/ART\`IS using an increasing number of CPU cores. A speedup lower than 1 means that the parallel simulation is slower than the sequential one. Conversely, a speedup larger than 1 is a gain for the parallel execution. Partitioning the simulated model in more CPU cores makes possible to share the model load. On the other hand, the communication between CPU cores (on the same die or on different dies) adds a communication cost (in terms of execution time).

For moderate loads (e.g.~1000-8000 SEs) the setup with 2 CPU cores is unable to obtain a speedup. This means that the communication cost added by the parallel setup is larger than the speedup given by parallelizing the model computation. For larger loads (e.g.~4000 up to 32000) there is always a gain. The general trend is that adding more CPU cores permits a better speedup. It is worth noting that the configuration with 32 CPU cores is obtained using all the logical cores provided by the Hyper-Threading technology while the real number of physical cores is 16. In both 16000 and 32000 SEs the best speedup is obtained when all the 16 physical cores are used (and in this case, Hyper-Threading is unable to give a speedup). For every SEs setup there is a parallel simulation that is able to reduce the WCT with respect to the sequential simulation. Despite of this, the speedup that can be obtained is not as good as expected. This is due to the characteristics of the simulated model: very little computation performed by each SE and a very high number of interactions among SEs. In this kind of model, with a such high level of interactions among SEs, it is not possible to expect a linear speedup when parallel/distributed execution architectures are used for running the simulations. Despite of this, we think that when the smart shire model will be more detailed (e.g.~implementing the specific behavior of each device in the smart shire) and with a better tuning of the communications in the simulated system, larger speedup values will be obtained.

Finally, the adaptive partitioning implemented in GAIA/ART\`IS has been enabled. The goal of this adaptive mechanism is to cluster the SEs in the CPU cores based on their interaction pattern. This means that, the SEs that are interacting with high frequency (e.g.~are in proximity in the simulated area) are clustered in the same CPU core. This, in most cases, permits a significant reduction of the communication cost in the parallel architecture. For example, because intra-process communication is faster than inter-process communication (especially if the processes are run on different CPU cores). Furthermore, computational load balancing strategies can be considered (e.g.~to deal with background load in the execution architecture).

\begin{figure}[h]
\centering
\includegraphics[width=8.5cm]{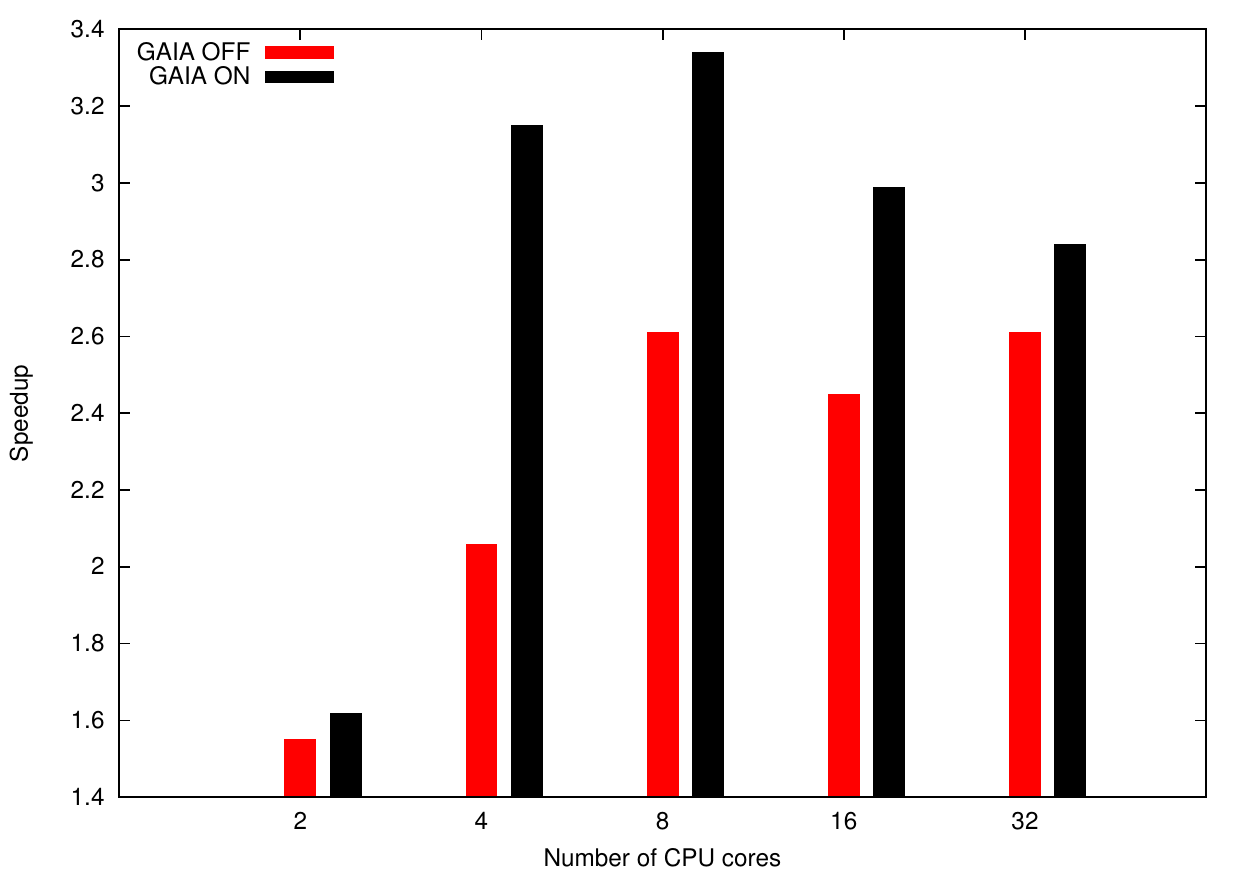}
\caption{Speedup of adaptive parallel simulation setup (different number of CPU cores, 32000 SEs).}
\label{fig_speedupon}
\end{figure}

In Figure~\ref{fig_speedupon} is reported the speedup than can be obtained for 32000 SEs with an increasing number of CPU cores and with the adaptive mechanism OFF (in red) or ON (in black). The adaptive mechanism always gives a gain with respect to static partitioning. This gain is very limited with 2 CPU cores but it becomes larger in all other cases. The best speedup is with 8 CPU cores, that is 3.34 vs. 2.61.  With 16 and 32 CPU cores there is still a gain but the overall speedup is lower. The rationale behind this behavior is that, up to a given number of CPU cores (i.e.~8) there is speedup that is due to the parallelization of the computational load in the simulated model. With a higher number of CPU cores, load parallelization given by the extra cores is balanced (or exceeded) by the extra cost for communication that is given by the larger parallel execution architecture. In the balanced case (e.g.~when 8 CPU cores are used) the computational load is properly partitioned and the adaptive mechanism for communication cost reduction is able to obtain a significant gain.

%%%%%%%%%%%%%%%%%%%%%%%%%%%%%%%%%%%%%%%%%%%%%%
\section{Conclusion}\label{sec:conc}
This paper discusses the need for novel solutions to foster the creation of smart services for countrysides. Such solutions must be cheap, adaptive, self-configuring and robust. Focusing on communication technologies, wireless solutions such as sensor networks, D2D and ad-hoc communications, multihoming techniques should be massively employed, in order to guarantee interconnection and content dissemination in areas with poor connectivity. Then, open data represents an important tool to be used during the development of smart services devised for smart shires. Finally, such services should be executed over scalable computing platforms such as cloud datacenters or fog computing infrastructures.

An important methodology to preliminarily assess the viability of large scale solutions in wide areas such as countrysides is simulation. In this case, simulation must be carefully handled in order to create reasonably accurate models that can scale in terms of modeled entities and granularity of events. In this sense, we demonstrated the viability of resorting to agent based simulation over a framework able to perform seamless sequential/parallel/distributed simulation.

In this work, we studied if message dissemination can be realized through a ``priority-based broadcast'' scheme over ad-hoc networks, where communication among nodes is made possible through direct transmissions among near devices. Results show that dissemination schemes (coupled with caching schemes) can represent an effective communication substrate to be used in a software middleware promoting the creation of applications for smart shire scenarios.

\small{
\bibliographystyle{abbrv}
\bibliography{paper}  
}

\end{document}